\documentclass[preprint,aps,epsfig,floatfix]{revtex4}
\usepackage{graphicx}% Include figure files
\usepackage{dcolumn}% Align table columns on decimal point
\usepackage{bm}% bold math
\usepackage{epsfig}
\usepackage{amsmath}
\begin{document}

\title{Effects of inhomogeneities and thermal fluctuations on the
  spectral function of a model d-wave superconductor}

\author{Daniel Valdez-Balderas} \email{valdez@pas.rochester.edu}

\affiliation{Department of Physics and Astronomy, University of Rochester, Rochester, New York, 14621}

\author{David Stroud} \email{stroud@mps.ohio-state.edu}

\affiliation{Department of Physics, The Ohio State University, Columbus,
  Ohio 43210}

\date{\today}

%-------------------------------------------------------------------
\begin{abstract}
  We compute the spectral function $A({\bf k}, \omega)$ of a model
  two-dimensional high-temperature superconductor, at both zero and
  finite temperatures $T$. The model consists of a two-dimensional BCS
  Hamiltonian with $d$-wave symmetry, which has a spatially varying,
  thermally fluctuating, complex gap $\Delta$.  Thermal fluctuations
  are governed by a Ginzburg-Landau free energy functional. We assume
  that an areal fraction $c_{\beta}$ of the superconductor has a large
  $\Delta$ ($\beta$ regions), while the rest has a smaller $\Delta$
  ($\alpha$ regions), both of which are randomly distributed in space.
  We find that $A({\bf k}, \omega)$ is most strongly affected by
  inhomogeneity near the point $\mathbf k = (\pi, 0)$ (and the
  symmetry-related points).  For $c_\beta\simeq 0.5$, $A({\bf k},
  \omega)$ exhibits two double peaks (at positive and negative energy)
  near this k-point if the difference between $\Delta_\alpha$ and
  $\Delta_\beta$ is sufficiently large in comparison to the hopping
  integral;
%REVISE: add phrase ``in comparison to the hopping integral''
  otherwise, it has only two broadened single peaks.
%  This double peak is consistent
%  with recent observations of a second branch of the dispersion
%  relation in underdoped LSCO using angle-resolved photoemission
%  spectroscopy [T.\ Yoshida {\it et al}, Phys.\ Rev.\ Lett.\ {\bf 91},
%  027001 (2003)].
   The strength of the inhomogeneity required to produce a split
   spectral function peak suggests that inhomogeneity is unlikely to
   be the cause of a second branch in the dispersion relation, such as
   has been reported in underdoped LSCO.
%REVISE: remove sentence involving Yoshida et al. and replace by the sentence immediately
%above. 
   Thermal fluctuations also affect $A({\bf k}, \omega)$ most strongly
  near $\mathbf k = (\pi,0)$.  Typically, peaks that are sharp at $T =
  0$ become reduced in height, broadened, and shifted toward lower
  energies with increasing $T$; the spectral weight near $\mathbf k =
  (\pi, 0)$ becomes substantial at zero energy for $T$ greater than
  the phase-ordering temperature.

\end{abstract}

\maketitle
%-------------------------------------------------------------------
\section{Introduction}

% for a good discussion of evidence of inhomogeneities from non-surface
% sensitive experiments, \ref{mayr_alvarez_moreo_dagotto} 
In the last several years, the measured electronic properties of
cuprate superconductors have shown considerable evidence of
inhomogeneities. For example, spatial variations of the
superconducting energy gap and of the local density of states spectrum
have been observed in scanning tunneling microscopy (STM) experiments
\cite{cren, howald1, pan_davis, lang_davis, howald2, fang,
kato_sakata, mashima}.  There have also been a number of reports of
magnetic and charge ordering in these materials, which also indicate
inhomogeneities ~\cite{cheong_aeppli, yamada, tranquada, niemoller,
mook, arai, pengcheng, hayden, sun}.  Other studies of cuprates have
shown that electronic states within certain energy ranges show
checkerboard-like spatial modulations~\cite{hanaguri_davis}.

A number of theoretical approaches have been developed to model these
inhomogeneities \cite{zaanen, emery_lin, emery_kivelson, low_emery,
emery_kivelson_tranquada, jamei, valdez_stroud1, martin_balatsky,
wang_pan, atkinson, nunner_andersen_hirscfeld, ghosal, cheng_su,
mayr_dagotto, jamei_kapitulnik_kivelson}.
%DANIEL, don't you need to cite around 12 articles here?
%DR. STROUD: done
These works are reviewed and extended in a recent article
\cite{valdez_stroud2}. In the present article, we use the approach of
Ref.\ \onlinecite{valdez_stroud2} to explore how the spectral function
of a d-wave superconductor is affected by gap inhomogeneities and
thermal fluctuations.

The spectral function of cuprate superconductors has been studied
theoretically by a number of groups, though most have omitted the
effects of quenched inhomogeneities.  For example, Wakabayashi {\it et
al.}\cite{wakabayashi_rice_sigrist} used a weak-coupling BCS theory
combined with a Green's function approach to explain the narrow
quasiparticle peak at the gap edge, which has been observed by ARPES
experiments in overdoped cuprates along the antinodal direction.
Pieri {\it et al.}\cite{pieri_pisani_strinati}, using a Nambu
formalism, have studied a model which includes pairing fluctuation
effects, and which accounts for some features of the single-particle
spectral function as observed in certain cuprates.  Zacher {\it et
al.}\cite{zacher_eder_arrigoni_hanke} have used a cluster perturbation
technique to compute the single-particle spectral function of the
$t$-$J$ and Hubbard models, and to study stripe phases in the
cuprates. Paramekanti {\it et al.}\cite{paramekanti_randeria_trivedi}
have studied a Hubbard model for (uniform) projected $d$-wave
states. They used a variational Monte Carlo technique in which one of
the variational parameters is the magnitude $\Delta$ of the pairing
field, and find that $\Delta$, as a function of doping, scales with
the
%DANIEL, with what feature of the hump?  The integrated area?  
%Also, what is T^\ast?
%DR. STROUD: In photoemission experiments on BSCCO below Tc [Campuzaon et al. 
%PRL 83, 3709 - 3712 (1999)], ARPES spectra show a peak followed at higher
%energies by a dip and then a hump. Both the variational parameter $\Delta$ 
%and the energy of the hump 
% scale with T* as function of doping.
% T* is the temperature at which the pseudogap opens as observed in 
% ARPES experiments
$(\pi,0)$ hump and $T^{\ast}$ as observed in ARPES experiments.

In another recent work, using a generalization of BCS theory, Chen
{\it et al.}  \cite{chen_levin_kosztin} found a sharpening of the
peaks in the spectral function as $T$ is reduced below $T_{c}$,
similar to what we find as discussed below.  (Their model involves a
homogeneous superconductor.)  Hotta et al.\cite{hotta_mayr_dagotto}
have used a self-consistent t-matrix approximation to study a model
for s and d wave superconductivity at finite $T$.  They found a gap in
both the single-particle density of states and the spectral function
even above the superconducting
%DANIEL, I assume it is the phase ordering transition
%
%DR. STROUD: In the summary of their article, Hotta et al mention that
%in their formalism it is very difficult to determine the
%superconducting transition temperature Tc, and instead they
%approximate it by a mean field transition temperature, which provies
%an upper limit to Tc. So I am not sure if for them Tc is a phase
%ordering temperature. Chen et al, on the other hand, mention in
%their conclusions that in their work emphasize the effects of
%long-range phase coherence which sets in at Tc. So I think in their
%case, yes, Tc is a phase ordering temperature.
transition temperature $T_c$; the energy scale for the pseudogap is
found to be the Cooper-pair binding energy.

More recently, Mayr {\it et al.}\cite{mayr_alvarez_moreo_dagotto} have
introduced an extended Hubbard model, which includes both
superconductivity and antiferromagnetism; they found that quenched
disorder is a necessary ingredient for that model to reproduce the
double branch, or split band, observed in angle-resolved photoemission
(ARPES) experiments on La$_{2-x}$Sr$_{x}$CuO$_{4}$.
%DANIEL, do they calculate the spectral function specifically?
%DR. STROUD: yes.
Finally, a model to study how $A({\bf k}, \omega)$ is affected by
thermal fluctuations of the phase, but not the amplitude, of the
superconducting order parameter has been treated by Eckl et
al.~\cite{eckl1} for homogenous systems.

In the present work, we propose a simple model for $A({\bf k},
\omega)$.  
%REVISE: The next sentence has been revised according to the referees' general suggestions.
This model can exhibit a split peak near $\mathbf k = (\pi, 0)$, but
only for certain parameter choices which are unlikely to be realized
experimentally.  Our model consists of a BCS superconductor with
$d$-wave symmetry, where the pairing field (given by the
superconducting order parameter) is inhomogeneous, and is also subject
to thermal amplitude and phase fluctuations at finite $T$.  We assume
that those thermal fluctuations are governed by a discretized
Ginzburg-Landau (GL) free energy functional.  To compute the spectral
function, we use exact numerical diagonalization of the BCS
Hamiltonian on a finite lattice and average over many different
configurations of the thermally fluctuating superconducting order
parameter as obtained using the Monte Carlo technique.  Thermal
averages are obtained by averaging over these configurations.

Inhomogeneities are introduced in our model phenomenologically.  The
atomic lattice is subdivided into cells, which we call XY cells, of
size $2 \times 2$ atomic sites; within each such cell, we assume
$\Delta$ to be constant.  Then we choose the coefficients of the GL
free energy functional so as to give, at $T = 0$, a binary
distribution of the superconducting order parameter $|\Delta|$ at each
atomic lattice site.  XY cells with small and large $|\Delta|$ values
are called $\alpha$ and $\beta$ cells, respectively.  We take the
distribution of $\alpha$ and $\beta$ cells on the atomic lattice to be
random, as suggested by STM experiments.  The two parameters which we
vary in our calculations are (i) the area fraction $c_\beta$ of
$\beta$ cells, and (ii) the magnitude $|\Delta_{\beta}|$ of the gap in
those cells.  The value of $|\Delta_{\alpha}|$ is kept the same
through our calculation, and is inferred from the STM experiments.
%DANIEL, I don't understand here - you changed \Delta_\beta but not \Delta?  Do
%you mean you changed \Delta_\beta but not \Delta_\alpha?
%DR. STROUD: yes, I corrected that sentence. 

At $T = 0$, we find that the main consequence of this binary, random
distribution of $\Delta$ is to broaden the peaks in the spectral
function $A({\bf k}, \omega)$ near the points $\mathbf k = (\pi,0)$
[and the symmetry-related points at $\mathbf k = (-\pi, 0)$ and $(0,
\pm \pi)$].
%where the gap has a node in the homogeneous d-wave
%superconductor]  
%REVISE: remove the last phrase beginning ``where'' since it is simply
%wrong: the gap has a node at \pi, \pi and symmetry-related points.
This broadening is most pronounced at $c_{\beta} = 0.5$.  For a
sufficiently large ratio $\Delta_\beta/\Delta_\alpha$ of the large to
the small gap, we find that $A({\bf k}, \omega)$ near $\mathbf k =
(\pi,0)$ shows two peaks rather than one (``split band regime'').
Otherwise, we find a single peak which is broadened by disorder.
Although the experimental ARPES results of Yoshida {\it et
  al.}~\cite{yoshida_uchida} also show a double peak, there are
several reasons to believe that this split peak is not caused by the
kind of inhomogeneities we consider here.  This point is discussed
further below.
%REVISE: Add last sentence above in response to the first referee.

At finite $T$, we find that, near $\mathbf k = (\pi, 0)$, the
originally sharp coherence peaks of $A({\bf k}, \omega)$ as a function
of $\omega$ broaden and shift to lower energies with increasing $T$.
%Furthermore, the gap starts filling up.
%DANIEL, by ``the gap starts filling,'' do you mean that A(k, \omega) 
%starts to become finite even at very low $\omega$?
%DR. STROUD: yes.
%DR. STROUD2: I commented out the sentence "Furthermore, the gap starts filling up" 
% which I had written before, I don't think it is necessary
%
This broadening is similar to that found in the calculations of Eckl
{\it et al.} \cite{eckl1}, which omits quenched disorder and also
include thermal fluctuations only in the phase but not the amplitude
of the gap.  These calculations focused on $T$ near that of the phase
ordering transition.  By contrast, we present calculations showing how
$A({\bf k}, \omega)$ evolves near $\mathbf k = (\pi, 0)$ as a function
of $\omega$ over a broad range of temperature, including both
amplitude fluctuations and quenched disorder.

The remainder of this paper is organized as follows.  In Section II,
we briefly describe our model, which is already presented in Ref.\
\onlinecite{valdez_stroud2}. In Section III, we give our numerical
results, followed by a discussion and conclusions in Section IV.

%-------------------------------------------------------------------
\section{Model}

\subsection{Microscopic Hamiltonian}
We consider the following Hamiltonian:
\begin{equation}
  H =   2 \sum_{\langle i,j \rangle,\sigma}t_{ij}c_{i\sigma}^{\dagger}c_{j\sigma}
+2\sum_{\langle i,j \rangle}(\Delta_{ij}c_{i\downarrow}c_{j\uparrow} +
\text{c.c.}) -\mu \sum_{i,\sigma}c_{i\sigma}^{\dagger}c_{i\sigma}
  \label{eq:hamil_bcs}
\end{equation}
Here, $\sum_{\langle i, j \rangle }$ denotes a sum over distinct pairs
of nearest neighbors on a square lattice with $N$ sites,
$c_{j\sigma}^{\dagger}$ creates an electron with spin $\sigma$
($\uparrow$ or $\downarrow$) at site $j$, $\mu$ is the chemical
potential, $\Delta_{ij}$ denotes the strength of the pairing
interaction between sites $i$ and $j$, and $t_{ij}$ is the hopping
energy, which we write as
\begin{equation}
  t_{ij} = -t_{hop}.  
  \label{eq:hopp_const}
\end{equation}
where $t_{hop} > 0$.

Following a similar approach to that of Ref.~\cite{valdez_stroud2} and
Ref.~\cite{eckl2} we take $\Delta_{ij}$ to be given by
\begin{equation}
  \Delta_{ij} = \frac{1}{4}\frac{|\Delta_i|+|\Delta_j|}{2} e^{i
  \theta_{ij}},
  \label{delta}
\end{equation}
where
\begin{equation}
  \theta_{ij} =
  \begin{cases}
    (\theta_i+\theta_j)/2, & \text{if bond $\langle i, j \rangle$ is in $x$-direction,}\\
    (\theta_i+\theta_j)/2 + \pi, & \text{if bond $\langle i, j \rangle$ is in $y$-direction,}
  \end{cases}
  \label{eq:thetaij}
\end{equation}
and
\begin{equation}
  \Delta_{j}=|\Delta_j|e^{i\theta_{j}},
  \label{eq:psi_complex}
\end{equation}
is the value of the complex superconducting order parameter at site
$j$.  The sums in~(\ref{eq:hamil_bcs}) are carried out over a lattice
we will refer to as the {\em atomic} lattice (as distinguished from
the {\em XY} lattice, described below).  The first term in
eq.~(\ref{eq:hamil_bcs}) corresponds to the kinetic energy, the second
term is a BCS type of pairing interaction with $d$-wave symmetry, and
the third term is the energy associated with the chemical potential.

%REVISE: Separate description of spectral function calculation into a special subsection.

\subsection{Numerical Calculation of Spectral Function}

We wish to compute the spectral function $A({\bf k}, \omega)$ for the
system described by the Hamiltonian (\ref{eq:hamil_bcs}).  Given
the $\Delta_{i}$'s, $t_{ij}$, and $\mu$, $A({\bf k}, \omega)$ is
computed through
\begin{equation}
  \text {A} (\omega,\mathbf k,\{ \Delta_{i} \} ) = \sum_{n, E_{n} \geq 0} [
  |u_{n}(\mathbf k)|^2 \delta(\omega-E_{n}) + |v_{n}(\mathbf k)|^2 \delta(\omega+E_{n}) ],
  \label{eq:spect_funct}
\end{equation}
where 
\begin{equation}
  u_{n}(\mathbf k) = \frac{1}{N^{1/2}} \sum_{i=1}^{N}\exp(i\mathbf k\cdot\mathbf r_i) u_{n}(\mathbf r_i),
  \label{eq:u_k}
\end{equation}
\begin{equation}
  v_{n}(\mathbf k) = \frac{1}{N^{1/2}} \sum_{i=1}^{N}\exp(i\mathbf k\cdot\mathbf r_i) v_{n}(\mathbf r_i),
  \label{eq:v_k}
\end{equation}
$E_n$ is the $n$th eigenenergy of Hamiltonian (\ref{eq:hamil_bcs}),
and
\begin{equation}
\psi_{n}(\mathbf r_{i}) = 
  \left[
    \begin{array}{c}
      u_{n}(\mathbf r_{i}) \\
      v_{n}(\mathbf r_{i})
    \end{array}
  \right],\quad\quad i=1,N.
%\psi_{n}(\mathbf r_i)\equiv[u_{n}(\mathbf r_i), v_{n}(\mathbf r_i)]^{t}
\end{equation}
is its $n$th eigenvector, as described in detail
in~\cite{valdez_stroud2}. Here,
\begin{equation}
  \mathbf r_{i} = a_{0} (n_{i}\hat x  + m_{i}\hat y ), 
\end{equation}
and $n_i$ and $m_i$ are integers in the range $[0,N_{x}-1]$ and
$[0,N_{y}-1]$.  
%REVISE: reword the next sentence to respond to comment (4) of second referee.
In our numerical calculations, we take the size of the atomic lattice
to be $N=N_{x}N_{y}$, where $\hat x$ and $\hat y$ are unit vectors in
the $x$ and $y$ directions, and $a_{0}$ is the lattice constant. We
use periodic boundary conditions, $\psi_{n}(\mathbf r) =
\psi_{n}(\mathbf r + N_{x} a_{0} \hat x)$ and $\psi_{n}(\mathbf r) =
\psi_{n}(\mathbf r + N_{y} a_{0} \hat y)$, which leads to $\mathbf
k$-vectors of the form
\begin{equation}
  \mathbf k = \frac{2 \pi}{a_{0}} \left( \frac{m_{x}}{N_{x}}\hat x + \frac{m_{y}}{N_{y}} \hat y \right)
\end{equation}

%REVISE: modify the paragraph below to respond to point (4) of the second referee.
The detailed procedure to obtain $\Delta_{i}$ is described in Ref.\ 
\onlinecite{valdez_stroud2}. Basically, we subdivide the atomic
lattice into cells, which we call $XY$ cells, of size $\xi_{0}\times
\xi_{0}$.  Here $\xi_{0}$ is the $T = 0$ Ginzburg-Landau (GL)
coherence length, which we take to be an integer multiple of $a_{0}$.
The value of $\Delta_{i}$ is assumed to be the same for each atomic
site within a given XY cell, and is governed by the following
discretized GL free energy functional:
\begin{equation}
\frac{F}{K_1} = \sum_{i=1}^M\left(\frac{T}{T_{c0i}} - 1\right)\frac{1}{\lambda_i^2(0)}\left|\frac{\Delta_i}
{k_BT_{c0i}}\right|^2 + \sum_{i=1}^M\frac{1}{18.76}\frac{1}{\lambda_i^2(0)}\left|\frac{\Delta_i}{k_BT_{c0i}}\right|^4
+ \sum_{\langle ij\rangle}\left|\frac{\Delta_i}{\lambda_i(0)k_BT_{c0i}} -
\frac{\Delta_j}{\lambda_j(0)k_BT_{c0j}}\right|^2.
\label{eq:gl}
\end{equation}
Here $K_1 = \hbar^4 d/[32(9.38)\pi m^{*,2}\mu_B^2]$, where $m^*= 2m_e$
is twice the electron mass, $\mu_B$ is the Bohr magneton, and $d$ is
the thickness of the superconducting layer.  If $d = 10\AA$, $K_1 =
2866$ eV $\AA^2$.  $\Delta_i$ is the complex gap parameter in the
i$^{th}$ XY cell.  In eq.\ (\ref{eq:gl}), the sums run over the
lattice of $XY$ cells, each of which contains $(\xi/a_0)^2$ atomic
sites.

We choose the coefficients of this GL free energy functional $T_{c0i}$
and $\lambda_i(0)$ to have binary distribution on the $XY$ lattice,
corresponding to either a small or a large value of $|\Delta_i|$.  We
call an $XY$ cell with a small (large) value of $|\Delta_{i}|$ an
$\alpha$ ($\beta$) cell, while the area fraction of $\beta$ cells is
called $c_{\beta}$.  The corresponding values of $T_{c0i}$ and
$\lambda_i(0)$ are denoted $T_{c0\alpha}$ and $T_{c0\beta}$.  At $T =
0$, in a homogeneous system made up entirely of $\alpha$ ($\beta$)
cells, the magnitude of $|\Delta_i|$ will be the same in each $XY$
cell and given by the minimum of the corresponding free energy
functional $F$, i.\ e.\ $|\Delta_i| = \sqrt{9.38}k_BT_{c0\alpha}$
($\sqrt{9.38}k_BT_{c0\beta}$).  In the binary case ($0 < c_\beta <
1$), at $T = 0$, we will still generally have $|\Delta_i| =
\sqrt{9.38}k_BT_{c0i}$, although this value may be modified slightly
by the proximity effect term in $F/K_1$ [the last term in eq.\ 
(\ref{eq:gl})].
%REVISE: Daniel, I added this entire paragraph.  Note the last sentence - do you think it is correct? 
% DR. STROUD2: yes.

We compute $A({\bf k}, \omega)$ at $T = 0$ by diagonalizing the model
hamiltonian (1) using $\Delta_i$ determined by minimizing the
Ginzburg-Landau free energy $F$.  This minimum value will always
correspond to gaps $\Delta_i = |\Delta_i|e^{i\theta_i}$ such that all
the phases $\theta_i$ are equal.
%REVISE: Daniel, is this correct (all phases being equal?)
%DR. STROUD2: yes
At finite $T$, we compute $A({\bf k},\omega)$ as an average over
different configurations $\{\Delta_i\}$.  These are obtained, as in
Ref.\ \cite{valdez_stroud2}, by assuming that the thermal fluctations
of the $\Delta_i$ are governed by the GL free energy functional $F$
described above.  Thus, $F$ is treated as an effective classical
Hamiltonian and thermal averages such as $\langle A({\bf k},
\omega)\rangle$ are computed as
\begin{equation}
\left\langle A({\bf k}, \omega)\right\rangle = \frac{\int\Pi_{i=1}^Nd^2\Delta_ie^{-F/k_BT}A({\bf k}, \omega,
\{\Delta_i\})}{\int\Pi_{i=1}^N d^2\Delta_ie^{-F/k_BT}}.
\end{equation}

% DR. STROUD3, added the paragraph below on Nov. 2, 2007, after
% receiving the second report from the referees.  It is essentially
% the same paragraph that you wrote on the vortex pinning
% article at the end of page 2 (page 2 as published in PRB)
We will be using the GL free energy functional at both $T = 0$ and
finite $T$ in spite of the fact that it was originally intended for
$T$ near the mean-field transition temperature. Strictly speaking, the
correct free energy functional near T = 0 should not have the GL form
but would be expected to contain additional terms, such as higher
powers of $|\psi|^2$. We use the GL form for convenience, and because
we expect that it will exhibit the qualitative behavior that would be
seen in a more accurate functional - that is, the effects of
inhomogeneities would be qualititatively the same in the GL model as
in a more accurate model containing additional powers of $|\psi|^2$.
%DANIEL: I added one long sentence to your paragraph - feel free to remove again if you
%think it is unnecessary.

To obtain $A({\bf k},\omega)$ for a given distribution of the
$\Delta_i$'s, we diagonalize the Hamiltonian (1) for that
configuration, then obtain $A({\bf k}, \omega)$ using eq.\ (6).  The
canonical averages are then evaluated using a Metropolis Monte Carlo
technique to determine the canonical distribution of the $\Delta_i$'s
at the temperature of interest.  The detailed description of this
Monte Carlo approach are given in Section IV.A of Ref.\ 
\cite{valdez_stroud2}.  As noted there, we first choose the values of
$T_{c0i}$ and $\lambda_i(0)$ in each $XY$ cell, taking these to be
quenched variables.  In contrast to our calculations of Ref.\ 
\cite{valdez_stroud2}, we do not include a smoothing magnetic field to
reduce finite-size effects; as a result, our results have more
numerical noise than do our earlier results for the density of states.

For the present model calculation, we arbitrarily set the chemical
potential $\mu=0$, for simplicity, and use $t_{hop}$ as the unit of
energy.  This corresponds to half filling in the band model.  Exactly
half filling would correspond to $x = 0$ in La$_x$Sr$_{1-x}$CuO$_4$
(LSCO), for example\cite{note2}.
%REVISE: Daniel, I did talk to Rajdeep and he confirmed that the above statement is correct.
It should be noted that some of the most interesting experimental
results for the spectral function\cite{yoshida_uchida} are carried out
in the underdoped superconducting regime of the phase diagram, where
$\mu$ is slightly negative.  If we set $\mu \neq 0$ in our model, this
leads to unequal integrated weights of the spectral function peaks at
positive and negative energy, but we have found that otherwise our
numerical results are not very different from those at
$\mu = 0$, for our model Hamiltonian.  However, the present results
and model, for reasons which we discuss below, are probably not
directly relevant to those experiments.

%REVISE: added the last two sentence above to emphasize that our model cannot be used to account for
%the results of Yoshida et al for the underdoped cuprates.  Should we include one figure in this regime?
%I am a bit reluctant to mention the antiferromagnetic regime at all, for fear that this will produce
%further misunderstandings by the referees.
%DR. STROUD2: I leave it up to you. I included the figure with mu = -0.05

%DR. STROUD2: I am adding the following tentative paragraph. Feel free to modify it,
% I didn't have much time to do the best writing. I think this could be merged with
% the paragraph above. I don't really mind if we include or not the equations that I
% wrote below.
In order to show that our results are not strongly affected by setting
$\mu = 0$, we have also done simulations using $\mu \neq 0$. For
example, in
Figure~\ref{plot_spect_funct_several_ktc_fact6_L12_nb800_pi_0_neg_mu}
we present results using $\mu = -0.05$. For this value of $\mu$, the
average number of electrons per site, defined as,
\begin{equation}
  \langle n \rangle = \frac{1}{N} \sum_{i=0}^{N} \langle n_{i} \rangle
\end{equation}
with
\begin{equation}
  n_{i}  = \sum_{\sigma} c_{i,\sigma}^{\dagger}c_{i,\sigma}
\end{equation}
is found to be $\langle n \rangle \sim 0.94$. This corresponds to a
strongly underdoped cuprate $x \sim 0.06$.

%The input to our problem is then a set of values of the coefficients
%of the GL free energy functional, in a discrete form. Relevant
%configurations of $\Delta_{i}$ are obtained with the use of Monte
%Carlo simulations. Once $\Delta_{i}$ at each atomic lattice site is
%known, the eigenvectors and eigenvalues of Hamiltonian
%(\ref{eq:hamil_bcs}) are obtained by exact diagonalization, and the
%spectral function is computed with the use of (\ref{eq:spect_funct}).

%In the present study, we set the GL coefficients so as to obtain, at
%zero temperature, either a homogeneous, or a binary distribution of
%$\Delta_{i}$ on the XY lattice. We call XY cells with a small value of
%$|\Delta_{i}|$ an $\alpha$ cell, while one with a large value is
%called $\beta$. The concentration of $\beta$ cells is called
%$c_{\beta}$.

\subsection{Homogeneous systems}

For a homogeneous system at $T = 0$, $\Delta_{i}=\Delta$ and we can rewrite
%REVISE: Daniel, I added the phrase ``at T = 0'' here.
Hamiltonian (\ref{eq:hamil_bcs}) as
\begin{equation}
  H =   \sum_{\mathbf{k},\sigma}\epsilon_{\mathbf{k}}c_{\mathbf{k}\sigma}^{\dagger}c_{\mathbf{k}\sigma}
+\sum_{\mathbf{k}}(\Delta_{\mathbf{k}}c_{\mathbf{k}\downarrow}c_{-\mathbf{k}\uparrow} +
\text{c.c.}) -\mu \sum_{\mathbf{k}}c_{\mathbf{k}\sigma}^{\dagger}c_{\mathbf{k}\sigma},
  \label{eq:hamil_bcs2}
\end{equation}
where 
%
%\begin{equation}
  $\epsilon_{\mathbf{k}} = -2 t [\cos(k_{x}a_{0})+\cos(k_{y}a_{0})]$
%\end{equation}
%
and
%
%\begin{equation}
  $\Delta_{\mathbf{k}} = \frac{1}{2}\Delta[\cos(k_{x}a_{0})-\cos(k_{y}a_{0})]$.
%\end{equation}
%
In obtaining (\ref{eq:hamil_bcs2}) we have used
%
%\begin{equation}
  $c_{j}^{\dagger} = \frac{1}{N^{1/2}}\sum_{\mathbf{k}}\exp(-i\mathbf k \cdot \mathbf r_{j})\,c_{\mathbf k}^{\dagger}$
%\end{equation}
%
and its hermitian conjugate.
%
%\begin{equation}
%  c_{j} = \frac{1}{N^{1/2}}\sum_{\mathbf{k}}\exp(i\mathbf k \cdot \mathbf r_{j})\,c_{\mathbf k}.
%\end{equation}
%
In this case, the excitation energies of the system are given by
\cite{tinkham}
\begin{equation}
  E_{\mathbf{k}} = \sqrt{(\epsilon_{\mathbf{k}} - \mu)^2 + \Delta_{\mathbf{k}}^{2} }
  \label{eq:excitationE}
\end{equation}
The corresponding spectral function will be a sum of two delta
functions, as indicated by eq.\ (\ref{eq:spect_funct}).
%DANIEL, is my last statement correct?  Also, I removed most of the 
%equation numbers, since this is standard, I believe.
%DR. STROUD: yes, your statement is correct.

\section{Numerical Results: Inhomogeneities and Thermal Fluctuations}

In this section we present our numerical results for $A({\bf k},
\omega)$ for inhomogeneous systems both at zero and finite
temperatures; for reference, we also show the corresponding results
for homogeneous systems in some cases.  For $T = 0$, we use $48 \times
48$ atomic lattices used, while at finite $T$ we used lattices of
$32\times 32$.  In al cases, we use $2 \times 2$ $XY$ cells.  Through
the rest of this article, we show energy measured in units of
$t_{hop}$, distance in units of $a_{0}$, and $\mathbf k$ in units of
$1/a_{0}$.

\subsection{Zero temperature}

%REVISE: add the paragraph below in response to point (2) of the
%second referee.

Before describing our results at zero temperature, we first comment
on our choice of gap parameters used in the calculations.  Our primary
goal is to ascertain what kinds of qualitative spectral functions could result
from the type of inhomogeneity described by our models, {\em not} to compare directly
to experiment.  For this reason, we will examine gaps which are, in general,
substantially larger (in units of $t_{hop}$) than those which would describe realistic
cuprate superconductors.   This point is examined further in the discussion
section.

\begin{figure}
  \setlength{\unitlength}{1.0in}
  \begin{picture}(4,6)%(12,7)% this pair of numbers seems to 'reserve' space. If it's too small, figure overlaps previous figure
    \put(0,2.9){
      \includegraphics*[width=2in, angle=0]{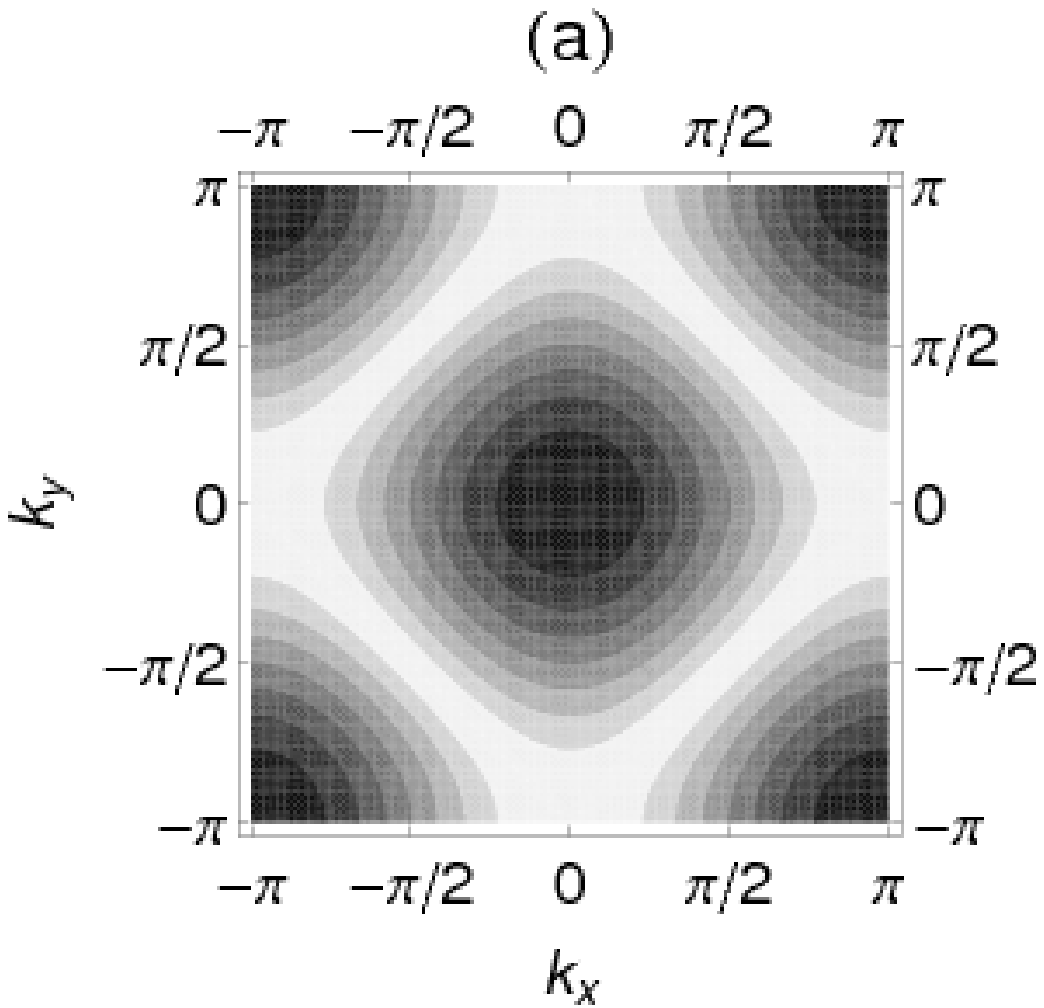}
    }
    \put(2.0,2.9){
      \includegraphics*[width=2in, angle=0]{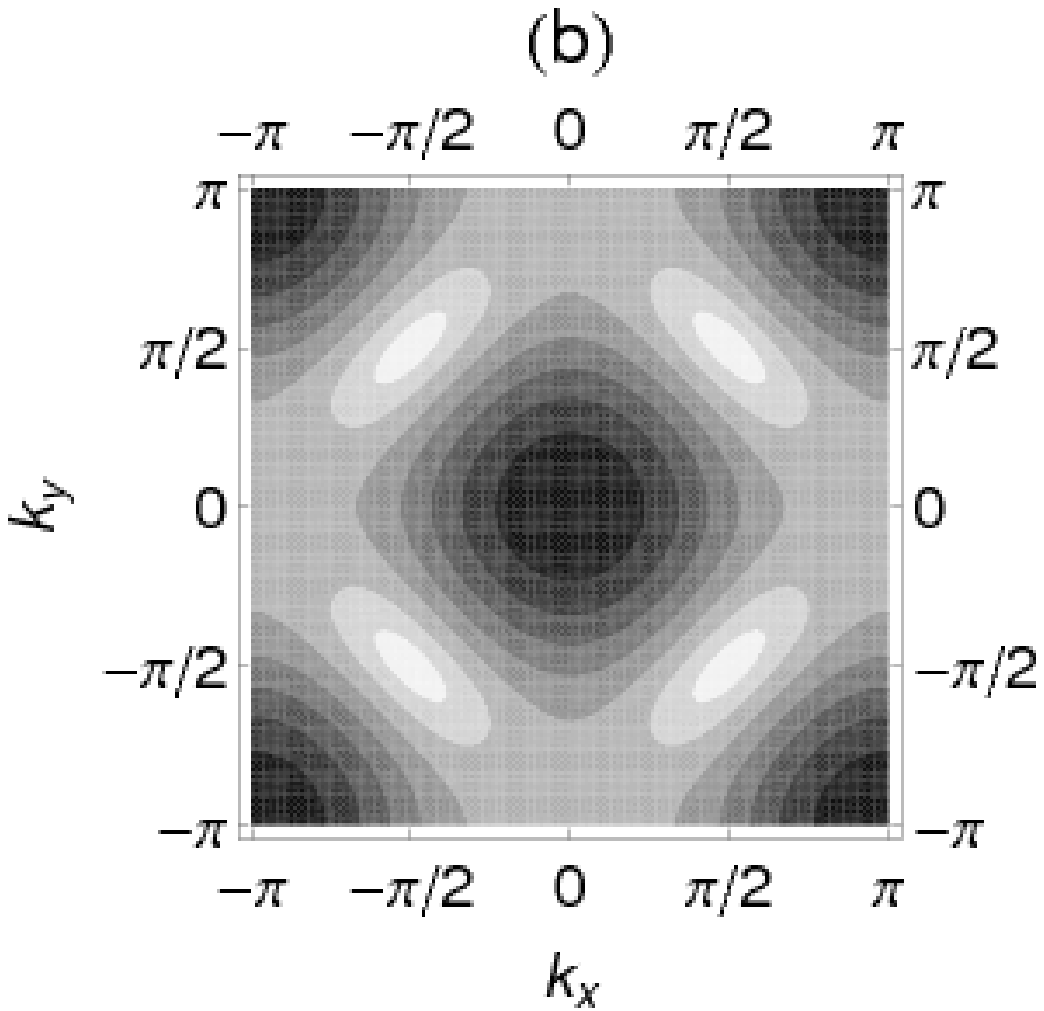}
    }
    \put(0,0){
      \includegraphics*[width=2in, angle=0]{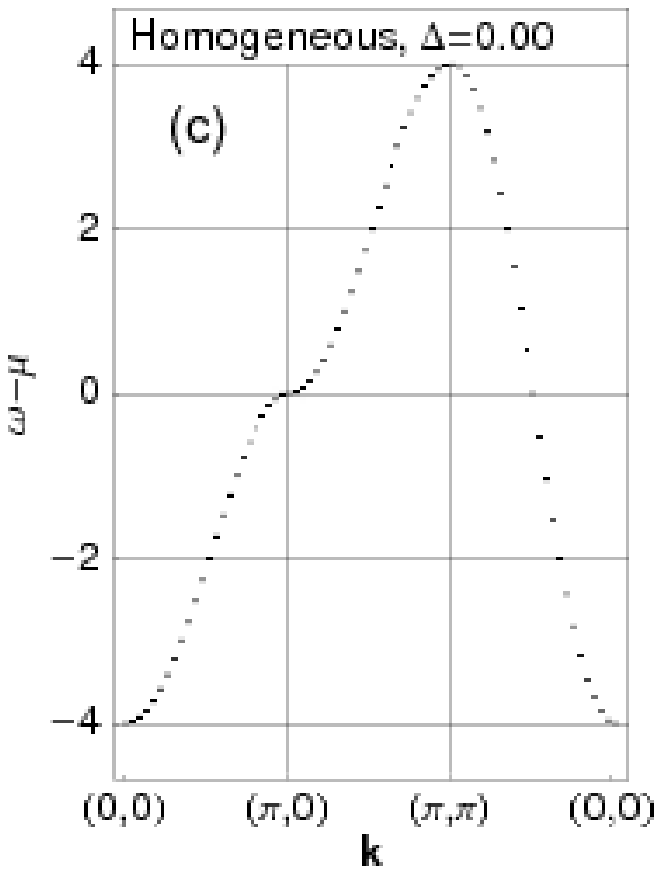}
    }
    \put(2.0,0){
      \includegraphics*[width=2in, angle=0]{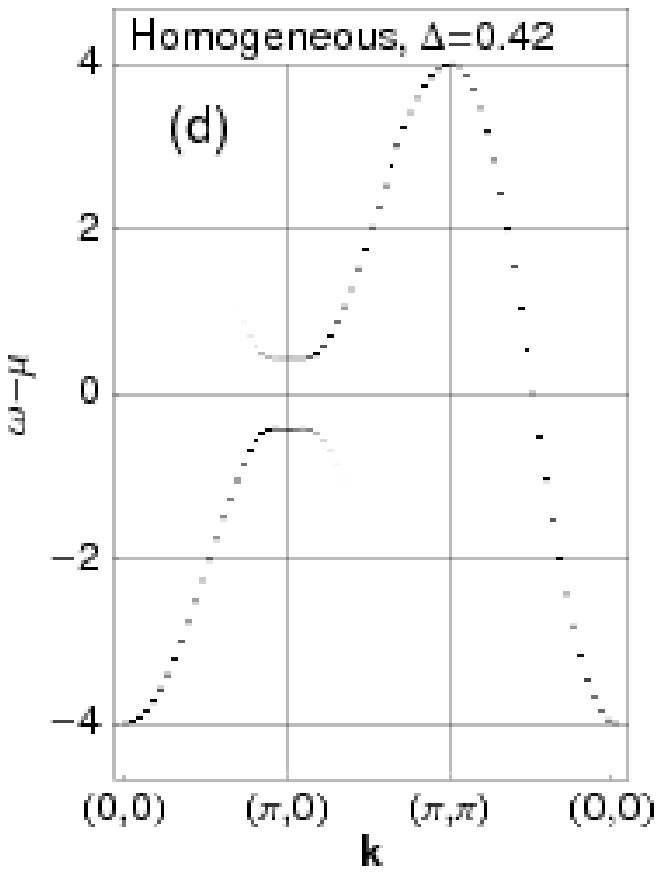}
    }

  \end{picture}
  \caption{\label{fig:pure}
    Contour plots of the energy $E_{\mathbf k}$ of single-particle
    excitations, and the corresponding spectral function spectral
    function $\text {A} (\mathbf k, \omega)$, for two homogeneous
    systems at zero temperature: one with $\Delta = 0$, and another
    with $\Delta = 0.42$.  In parts (a) and (b), the dark (light)
    regions correspond to large (small) values of the excitation
    energies, as calculated from Eq.~(\ref{eq:excitationE}).  In parts
    (c) and (d), we plot the positions of the peaks of $\text {A}
    (\mathbf k, \omega)$ for these two homogeneous systems.}
\end{figure}

With this preamble, we now present our results at $T = 0$.  Fig.~\ref{fig:pure} shows
the spectral function $A({\bf k}, \omega)$ (represented as a contour
plot) as well as plots of the dispersion relation $E_{\bf k}$ as a
function of ${\bf k}$, for two homogeneous systems: one with $\Delta =
0$, and another with $\Delta = 0.42$.   For such homogeneous systems,
$A({\bf k}, \omega)$ is simply proportional to the sum of two delta functions:
$A({\bf k}, \omega) \propto \delta(\omega - E_{\bf k}) + \delta(\omega + E_{\bf k})$.  
%REVISE: rewrite A(k, \omega) to conform to comment (5) of the second Referee that
%the spectral function for a homogeneous system is the sum of two delta functions.  
%Daniel, is this correct as written?
%DR. STROUD2: I didn't have time to see if it is correct, sorry
In parts (a) and (b), the dark (light)
regions correspond to regions where $E_{\bf k}$, as calculated from
Eq.~(\ref{eq:excitationE}), is large (small); these are shown for all
$\mathbf k$ vectors in the first Brillouin zone (BZ). For a system
with $\Delta =0$ [Fig.~\ref{fig:pure}(a)] there are four lines (white)
in $\mathbf k$-space for which $E_{\mathbf k} = 0$: $k_{y}=\pm
k_{x}\pm \pi$.  When $\Delta > 0$ [Fig.~\ref{fig:pure}(b)], the lines
are reduced to four points: ($k_x=\pi / 2$, $k_y=\pm \pi /2$) and
($k_x=-\pi / 2$, $k_y=\pm \pi /2$), located at the center of the white
blobs in Fig.~\ref{fig:pure}(b).
In Fig.~\ref{fig:pure}(c) and (d), density plots of $\text {A}
(\mathbf k, \omega)$ as a function of $\omega$ are presented for those
homogeneous systems at selected $\mathbf k$ values. These values lie
along three standard lines in the first BZ: from $\mathbf k = (0,0)$
to $\mathbf k = (\pi,0)$, from $\mathbf k = (\pi,0)$ to $\mathbf k =
(\pi,\pi)$, and from $\mathbf k = (\pi,\pi)$ to $\mathbf k = (0,0)$.
Dark (light) regions correspond to large (small) values of the
spectral function.  For each ${\bf k}$ in these homogeneous systems,
there is a sharp peak in $A({\bf k}, \omega)$, whose energy and width
are indicated as the very short dashed lines in the plot.  Also, the
spectral function is clearly most strongly affected by a finite value
of $\Delta$ near $\mathbf k = (\pi, 0)$, where an energy gap of
magnitude $\Delta$ opens around $\omega - \mu = 0$.
%DANIEL, what is \mu here?  Is it zero?
%DR. STROUD: yes, \mu is zero throughout the article. 

\begin{figure}
  \setlength{\unitlength}{1.0in}
  \begin{picture}(4.5,6)%(12,7)% this pair of numbers seems to 'reserve' space. If it's too small, figure overlaps previous figure
    \put(0,2.7){
      \includegraphics*[width=2in, angle=0]{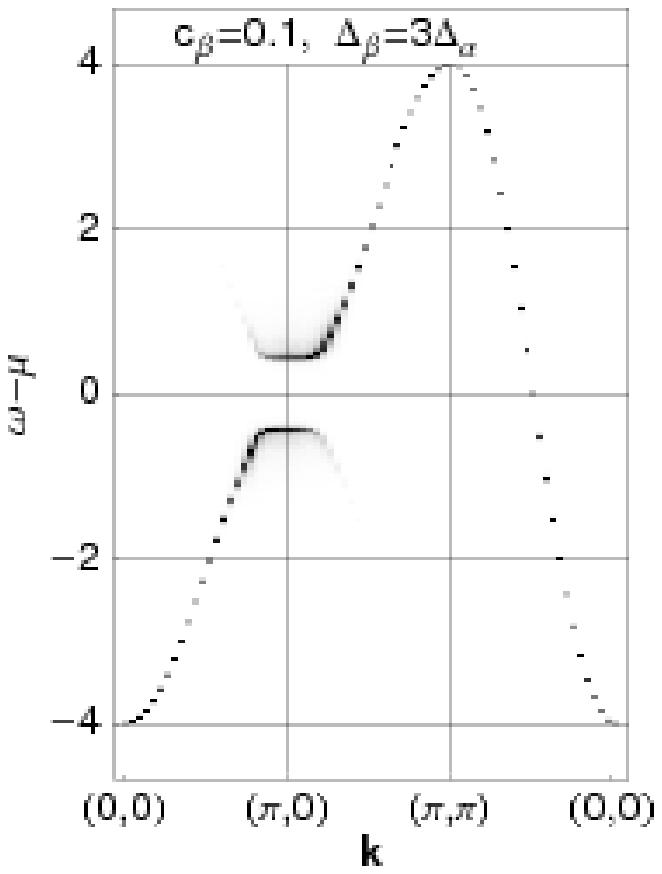}
    }
    \put(2.0,2.7){
      \includegraphics*[width=2in, angle=0]{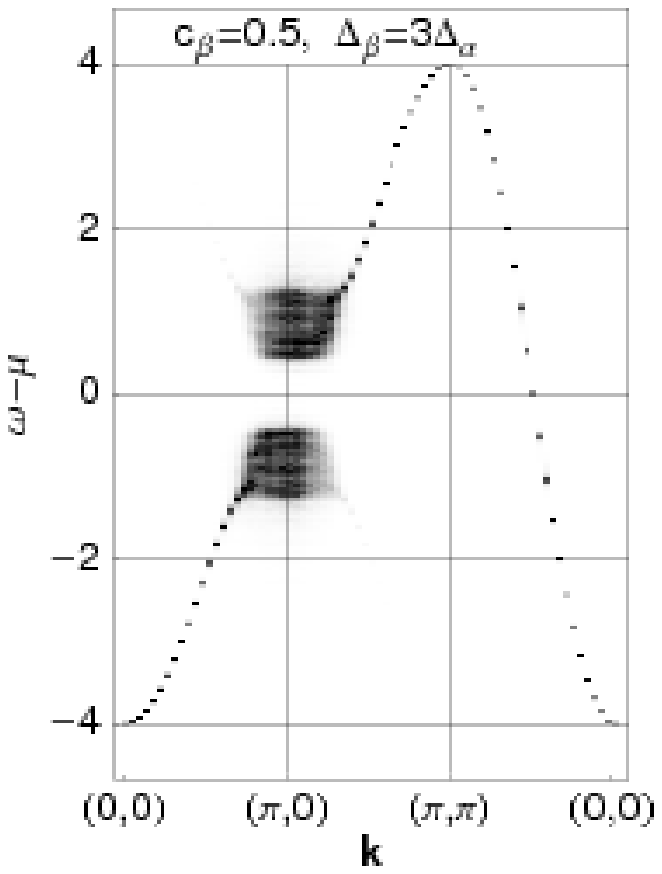}
    }
    \put(0,0){
      \includegraphics*[width=2in, angle=0]{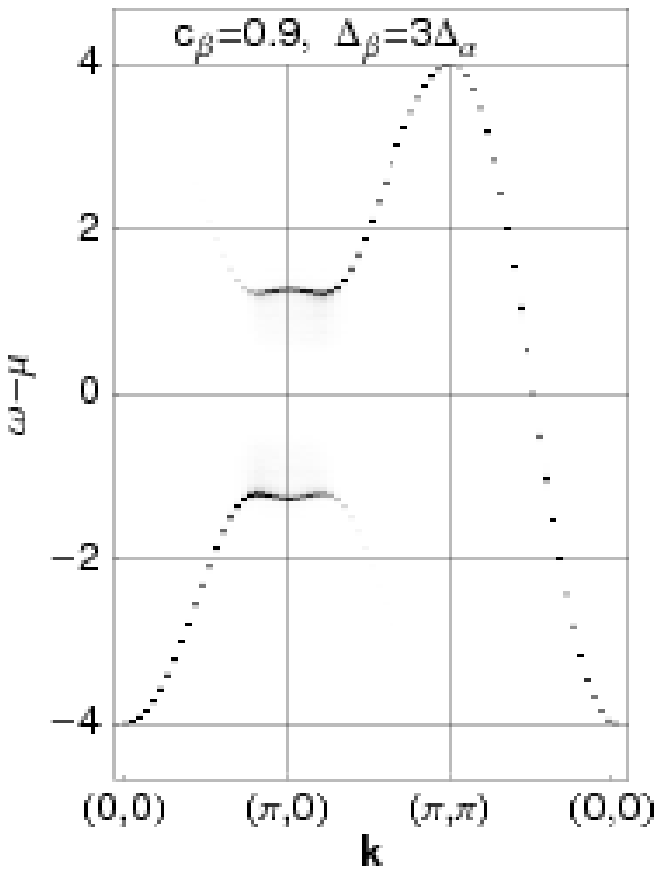}
    }
    \put(2.0,0){
      \includegraphics*[width=2in, angle=0]{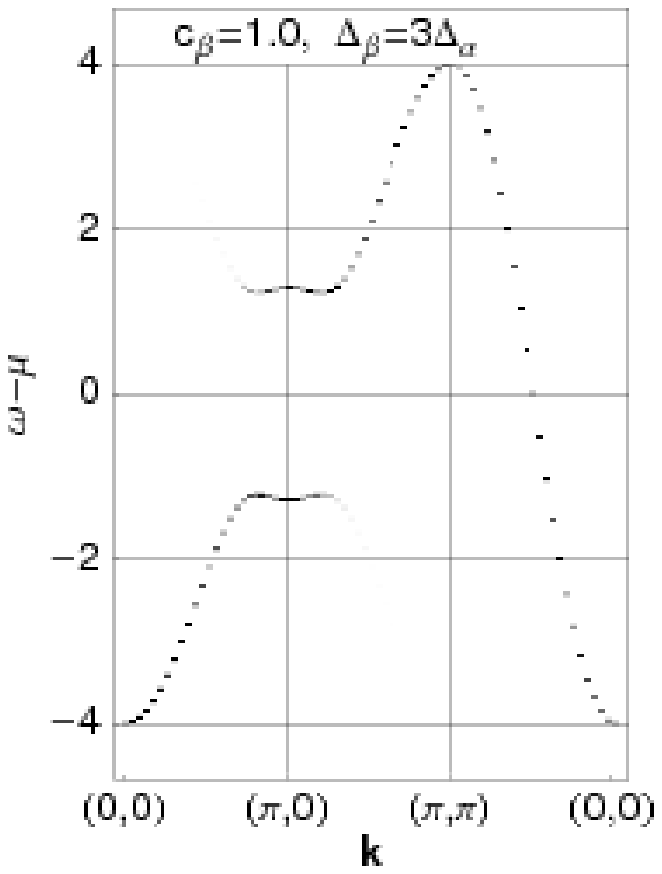}
    }
  \end{picture}
  \caption{\label{fig:mixed_fact3}
    Same as Fig.\ 1(c) and 1(d), but for several inhomogeneous systems, with
    different concentrations of $\beta$ (large-gap) cells, at
    $T = 0$.  In all four plots, atomic sites within the 
    $\alpha$ and $\beta$ cells have $\Delta = 0.42$ and $\Delta = 1.26$, respectively;
    the cells are randomly distributed over the atomic lattice, as illustrated in
    Fig.~\ref{fig:lattice_dis} for the case
    $c_{\beta}=0.1$.   The dark (light) regions correspond to large (small)
    values of $\text {A} (\mathbf k, \omega)$.  A more quantitative view of
    $\text {A} (\mathbf k, \omega)$  is shown in
    Fig.~\ref{fig:multiplot_spect_funct_fact3_1.0pi_0} for $\mathbf k = (\pi, 0)$.
%    
%    In $\beta$ cells, $\Delta = 1.26$, while in $\alpha$ cells $\Delta
%    = 0.42$. $\beta$ cells are randomly distributed over the atomic
%    lattice, as is illustrated in Fig.~\ref{fig:lattice_dis} for the
%    case $c_{\beta} = 0.1.$
}
\end{figure}

\begin{figure}[ht]\centering
  {\includegraphics[width=8cm]{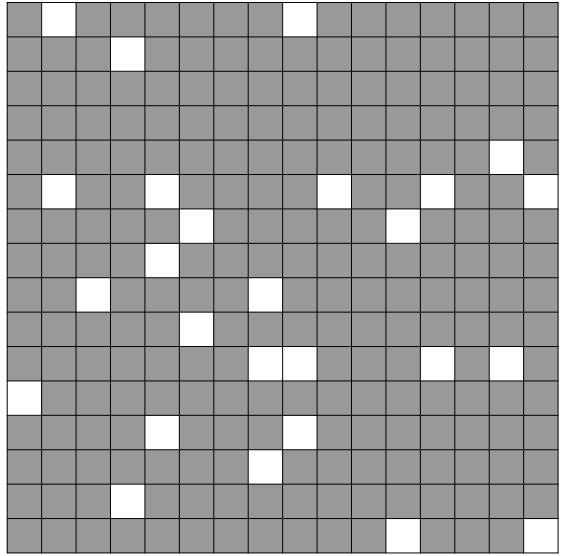}}
  \caption{
    A typical realization of disorder in a system with a
    concentration $c_{\beta} = 0.1$ of $\beta$ cells (white) immersed
    in a background of $\alpha$ cells (grey). Each cell contains four
    atomic sites.}
  \label{fig:lattice_dis}
\end{figure}

% changes to proof 
Figures~\ref{fig:mixed_fact3} and 3 show the spectral function of several inhomogeneous systems with
different concentrations $c_{\beta}$ of $\beta$ cells, at $T = 0$.  In
these systems, the atomic cells within the $\beta$ cells have $\Delta
= 1.26$, and are randomly distributed in the atomic lattice, while
$\alpha$ cells, which occupy the rest of the lattice, have $\Delta 
0.42$.  Fig.~\ref{fig:lattice_dis} shows a representative arrangements
of $\alpha$ and $\beta$ cells for an $16 \times 16$ XY lattice with
$c_{\beta}=0.1$.
We can observe in Fig.~\ref{fig:mixed_fact3} that the disorder
introduced by this binary distribution of the superconducting order
parameter affects mostly the region $\mathbf k = (\pi,0)$. This
disorder effect is almost unobservable for $c_{\beta} = 0.9$: the
results are almost the same as those for $c_\beta = 1.0$, a
homogeneous system with only $\beta$ cells.  On the other hand, a
small but noticeable disorder effect is observed $c_{\beta} = 0.1$, in
the form of a slight broadening of the spectral function at $\mathbf k
\simeq (0.8\pi,0)$.  However, it is the system with $c_{\beta} = 0.5$
the one that shows a most dramatic blurring of the energy in the
region of $\mathbf k = (\pi, 0)$, as we now discuss.

\begin{figure}[ht]\centering
  {\includegraphics[width=10cm]{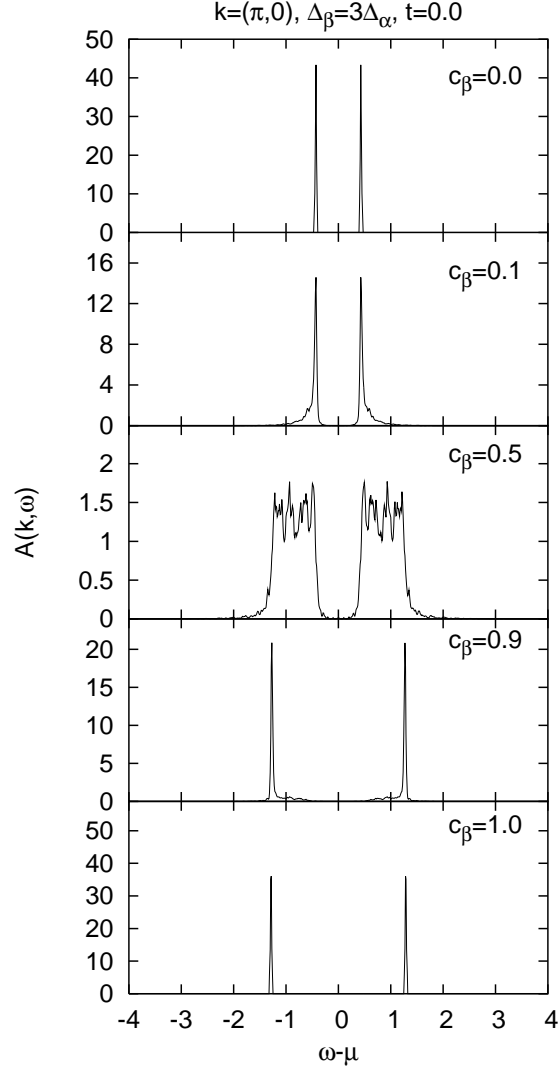}}
  \caption{
    Spectral function $\text {A} [\mathbf k = (\pi,0), \omega]$ as a
    function of $\omega$ of systems with different concentrations
    $c_{\beta}$ of $\beta$ cells at zero temperature.  We have taken
    $\Delta_\alpha = 0.42 t_{hop}$, $\Delta_\beta = 1.26 t_{hop}$}
%REVISE: added specification of parameters used in this figure.    
  \label{fig:multiplot_spect_funct_fact3_1.0pi_0}
\end{figure}

%REVISE: In the next two paragraphs, I tried to correct several misprints,
%including those mentioned by Referee 2 in his point (4) and several others
%which I saw.  
Since the effects of the binary distribution of $\Delta$ are more
pronounced near $\mathbf k = (\pi, 0)$, we have also plotted $\text
{A} (\mathbf k, \omega)$ versus $\omega$ for fixed $\mathbf k = (\pi,
0)$, at different values of $c _{\beta}$ in
Fig.~\ref{fig:multiplot_spect_funct_fact3_1.0pi_0}. For the pure
$\alpha$ system, $c_{\beta}=0.0$, two sharp peaks appear at $|\omega |
= \Delta_{\alpha}= 0.42$. When a fraction 0.1 of the $\alpha$ XY cells are
replaced by $\beta$ cells, $c_{\beta}=0.1$, the height of the peaks
decreases from about 45 (arbitrary units) to about 15, with a
corresponding broadening of the peak and a shifting of the weight
toward a higher energy. At $c_{\beta}=0.5$, the peak height is only
about 1.5, and the width is very large; the peak fills the entire
frequency range from $\omega = \Delta_{\alpha}=0.42$ to
$\Delta_{\beta}=1.26$. At $c_{\beta}=0.9$, most of the weight of
$\text {A} (\mathbf k, \omega)$ shifts to $\omega =
\Delta_{\beta}=1.26$, with a slight broadening near the bottom of the
peaks, which is, however, less pronounced than the corresponding
broadening of the $c_{\beta}=0.1$ peaks.  At $c_{\beta}=1.0$ the peaks
become sharp at $\omega = \pm \Delta_{\beta}= \pm 1.26$.

\begin{figure}[ht]\centering
  {\includegraphics[width=10cm]{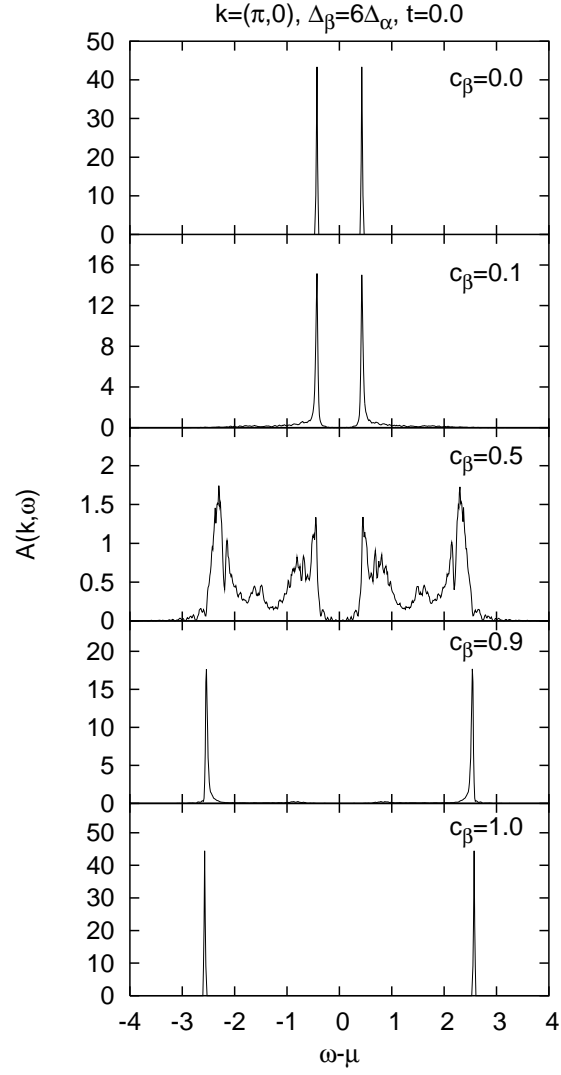}}
  \caption{
    Same as Fig.~\ref{fig:multiplot_spect_funct_fact3_1.0pi_0} but for
    $\beta$ cells which have $\Delta = 2.52$ instead of $\Delta =
    1.26$.}
  \label{fig:multiplot_spect_funct_fact6_1.0pi_0}
\end{figure}

We now discuss the calculated effects of a binary gap distribution on
systems similar to those of Fig.~\ref{fig:mixed_fact3}, but with
$\Delta_{\beta}= 2.52$ instead of $\Delta_{\beta}= 1.26$. These
systems, like the one previously discussed, have $\Delta_{\alpha}=
0.42$. For concentrations $c_{\beta}=0.1$ and $c_{\beta}=0.9$, the
effect of inhomogeneities qualitatively resembles that seen for
$\Delta_{\beta}= 1.26$: they produce broadening of the spectral
function near $\mathbf k = (\pi, 0)$. The main difference is that the
broadening is slightly greater for $\Delta_\beta = 2.52$.  However,
the case $c_{\beta}=0.5$ shows a real qualitative change: the spectral
function now splits into two well-defined peaks for $\mathbf k$
vectors near $(\pi,0)$.  We can better visualize this effect by
looking at Fig.~\ref{fig:multiplot_spect_funct_fact6_1.0pi_0}, where
we plot $\text {A} (\mathbf k, \omega)$ versus $\omega$ for fixed
$\mathbf k = (\pi, 0)$ and several values of $c_{\beta}$.  Clearly,
$\text {A} (\mathbf k, \omega)$ for $c_{\beta}=0.1$ and $c_{\beta}=0.9$
behaves similarly to the case $\Delta_{\beta}= 1.26$: slightly
broadened peaks at an energy near the $\Delta$ of the majority of the
XY cells, i. e., at $\omega = 0.42$ for $c_{\beta}=0.1$ and at $\omega
= 2.52$ for $c_{\beta}=0.9$.  But for $c_{\beta}=0.5$, $\text
{A} (\mathbf k, \omega)$ shows several peaks, two of which are
particularly clear: one at $\omega \simeq 0.42$ and the other at
$\omega \simeq 2.52$.  This is the ``split band'' regime one expects
for large contrast between $\Delta_\alpha$ and $\Delta_\beta$.

%REVISE: add the paragraph below and add one additional Figure (the one Daniel
%plotted showing the evolution of the split band regime as a function of gap
%magnitude, for fixed $|\Delta_\beta/\Delta_\alpha| = 6.$

\begin{figure}[ht]\centering
{\includegraphics[width=10cm]{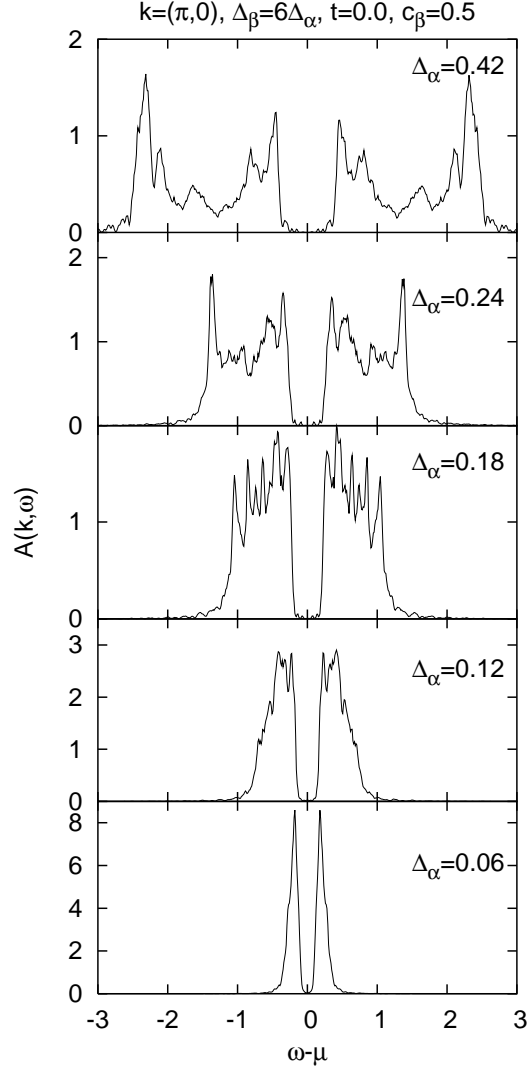}}
\caption{
Spectral function $\text{A}[\mathbf k = (\pi, 0), \omega]$ as a function
of $\omega$ for a system with concentration $c_\beta = 0.5$ of $\beta$ cells at $T = 0$,
plotted as a function of the magnitude $|\Delta_\alpha|$ of the component with
the smaller gap.  In all cases, $|\Delta_\beta/\Delta_\alpha| = 6$.}
\label{fig:mixed_fact6new}
\end{figure}

\begin{figure}[ht]\centering
{\includegraphics[width=10cm]{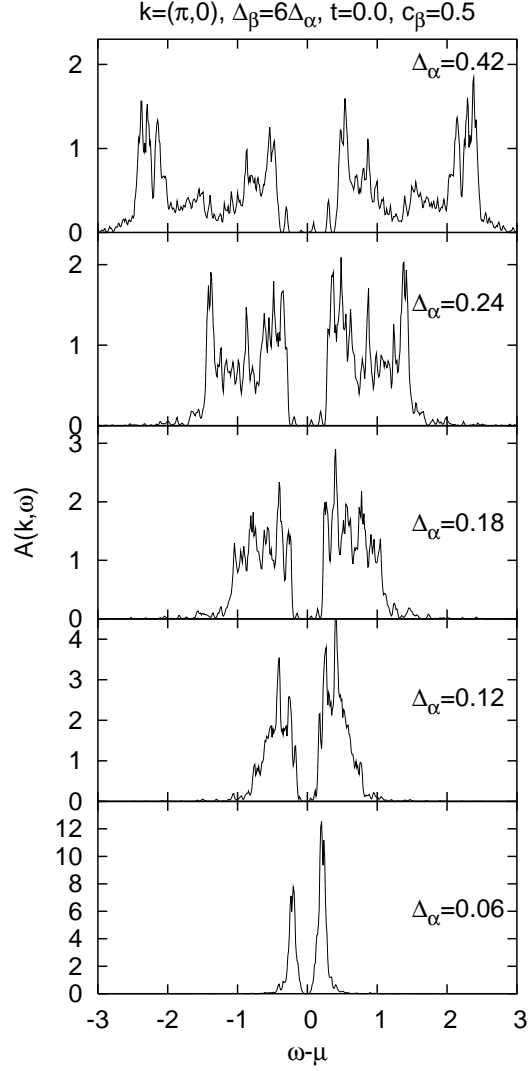}}
\caption{
  Same as Fig.~\ref{fig:mixed_fact6new}, but with chemical potential
  $\mu = -0.05$ instead of $\mu = 0$. With our model, setting $\mu =
  -0.05$ results in having an average occupation number $\langle n
  \rangle \sim 0.94$, which
  corresponds to an strongly underdoped cuprate $x \sim 0.06$.}
\label{plot_spect_funct_several_ktc_fact6_L12_nb800_pi_0_neg_mu}
\end{figure}

In order to better visualize how the spectral function depends on disorder, 
we have calculated $A({\bf k}, \omega)$ as function of $|\Delta_\alpha|$ for a
fixed ratio $|\Delta_\beta/\Delta_\alpha| = 6$, at $c_\beta = 0.5$.  The results
are shown in Figs.\ \ref{fig:mixed_fact6new} and 8.  This series of plots clearly shows
the evolution of $A({\bf k}, \omega)$ from a split-band regime at $|\Delta_\beta| = 2.52$
(in units of $t_{hop}$) to a broadened single band for $|\Delta_\beta| = 0.22 |t_{hop})$ or
smaller.
%REVISE: Daniel, aren't the Delta's measured in units of t_\hop rather than in meV?
%DR. STROUD2: yes, all energies, including Delta's, are measured in units of t_{hop}
In general, we find that the split band regime occurs only if the difference 
$|\Delta_\beta|-|\Delta_\alpha|$ is of order $t_{hop}$ or larger; otherwise, $A({\bf k}, \omega)$
at ${\bf k} = (\pi, 0)$ is simply the sum of two broadened peaks at positive and negative
energies.

%==================================================================

\subsection{Finite temperatures}

\begin{figure}
  \setlength{\unitlength}{1.0in}
  \begin{picture}(4.5,6)%(12,7)% this pair of numbers seems to 'reserve' space. If it's too small, figure overlaps previous figure
    \put(0,2.7){
      \includegraphics*[width=2in, angle=0]{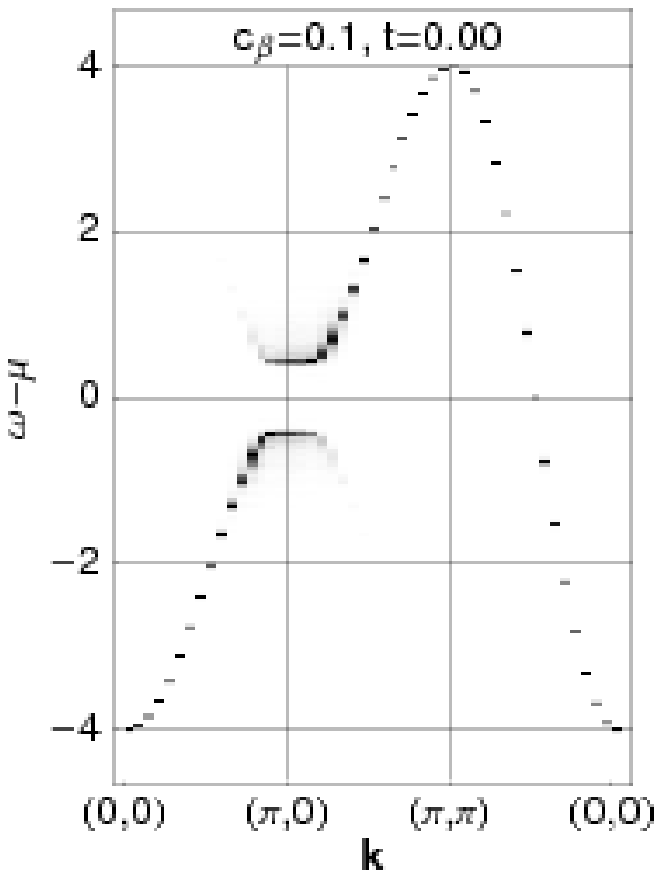}
    }
    \put(2.0,2.7){
      \includegraphics*[width=2in, angle=0]{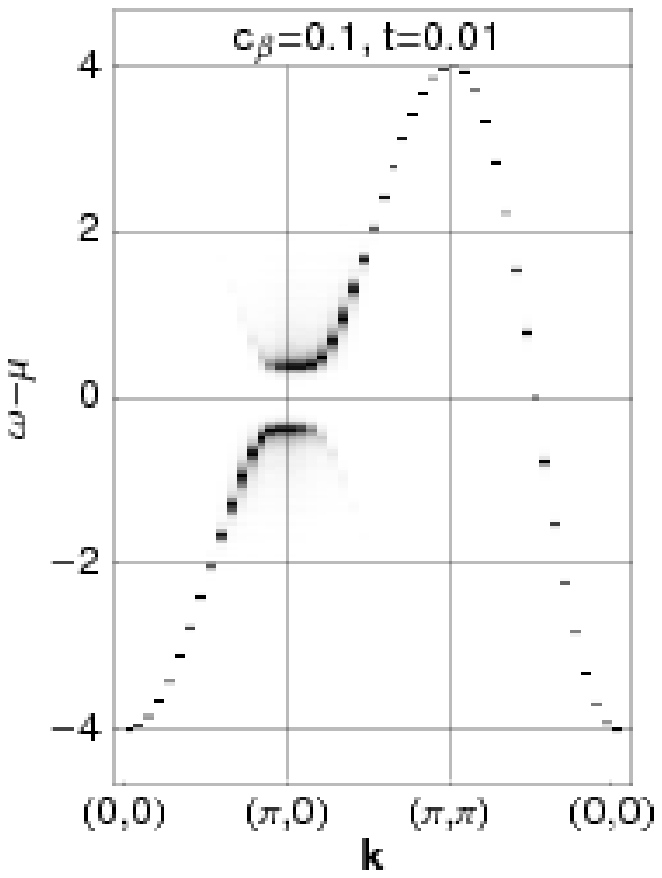}
    }
    \put(0,0){
      \includegraphics*[width=2in, angle=0]{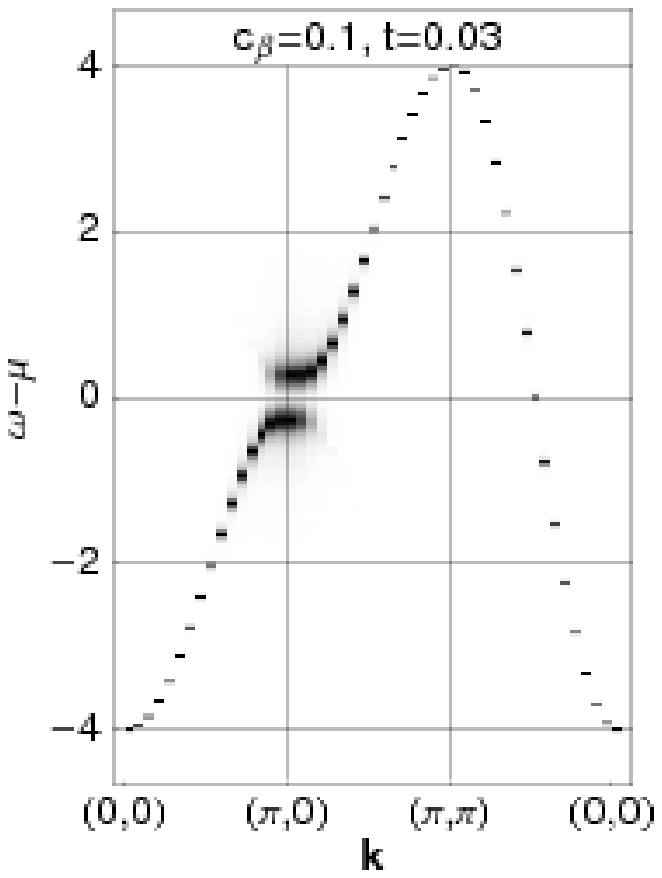}
    }
    \put(2.0,0){
      \includegraphics*[width=2in, angle=0]{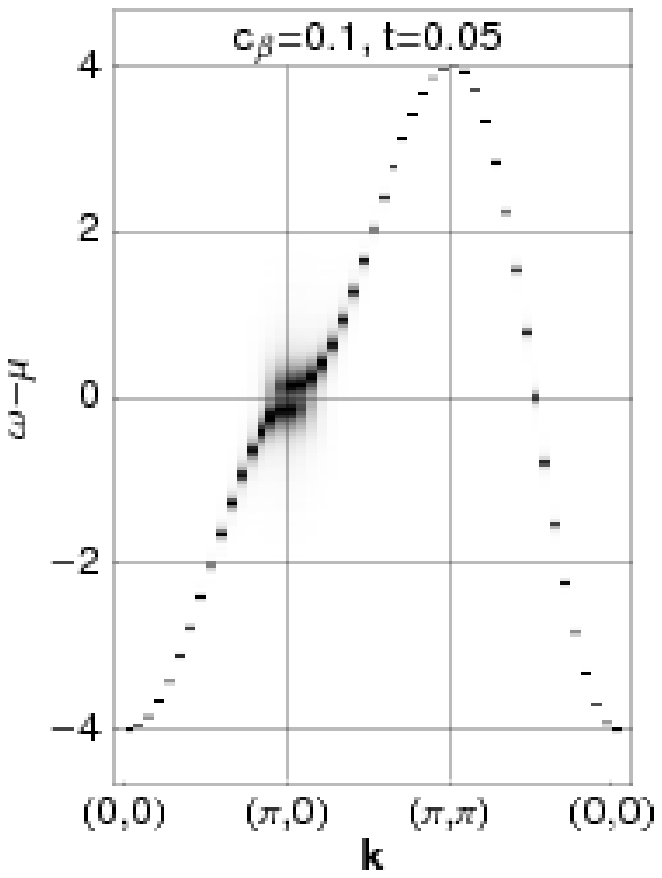}
    }
  \end{picture}
  \caption{\label{fig:mixed_finite_temp}
    Plots of the spectral function$\text {A} (\mathbf k, \omega)$ for a system at
    zero and finite temperatures.  The system has a concentration
    $c_{\beta}=0.1$ of $\beta$ cells with $\Delta_\beta = 1.26$ randomly
    distributed in a background of $\alpha$ cells having
    $\Delta_\alpha = 0.42$. The phase-ordering temperature is $t_{c}\approx
    0.035$.  The detailed evolution of the curve $\text {A}
    [\mathbf k = (\pi,0), \omega]$ versus $\omega$ can be seen in
    Fig.~\ref{fig:multiplot_spect_funct_allT_c0.1}.}
\end{figure}

Fig.~\ref{fig:mixed_finite_temp} shows the $T$-dependence of $A({\bf
  k},\omega)$ for a system with $c_{\beta}=0.1$.  The $\alpha$ and
$\beta$ cells are now characterized by values of $t_{c0}$ such that at
low $T$, $\Delta = 0.42$ in the $\alpha$ cells, and $1.26$ in the
$\beta$ cells. The value of $\Delta$ itself at finite $T$ will, of
course, thermally fluctuate, as governed by the GL free energy
functional discussed at the end of Section II.  The spectral function
presented here is therefore an average of $\text {A} (\omega,\mathbf
k,\{ \Delta_{i} \} )$ over different configurations $\{ \Delta_{i} \}$
obtained by a Monte Carlo sampling procedure, as described above
and in Ref.\ \cite{valdez_stroud2}.  Hereafter, we denote this ensemble
average simply as $A({\bf k}, \omega)$.

\begin{figure}[ht]\centering
  {\includegraphics[width=10cm]{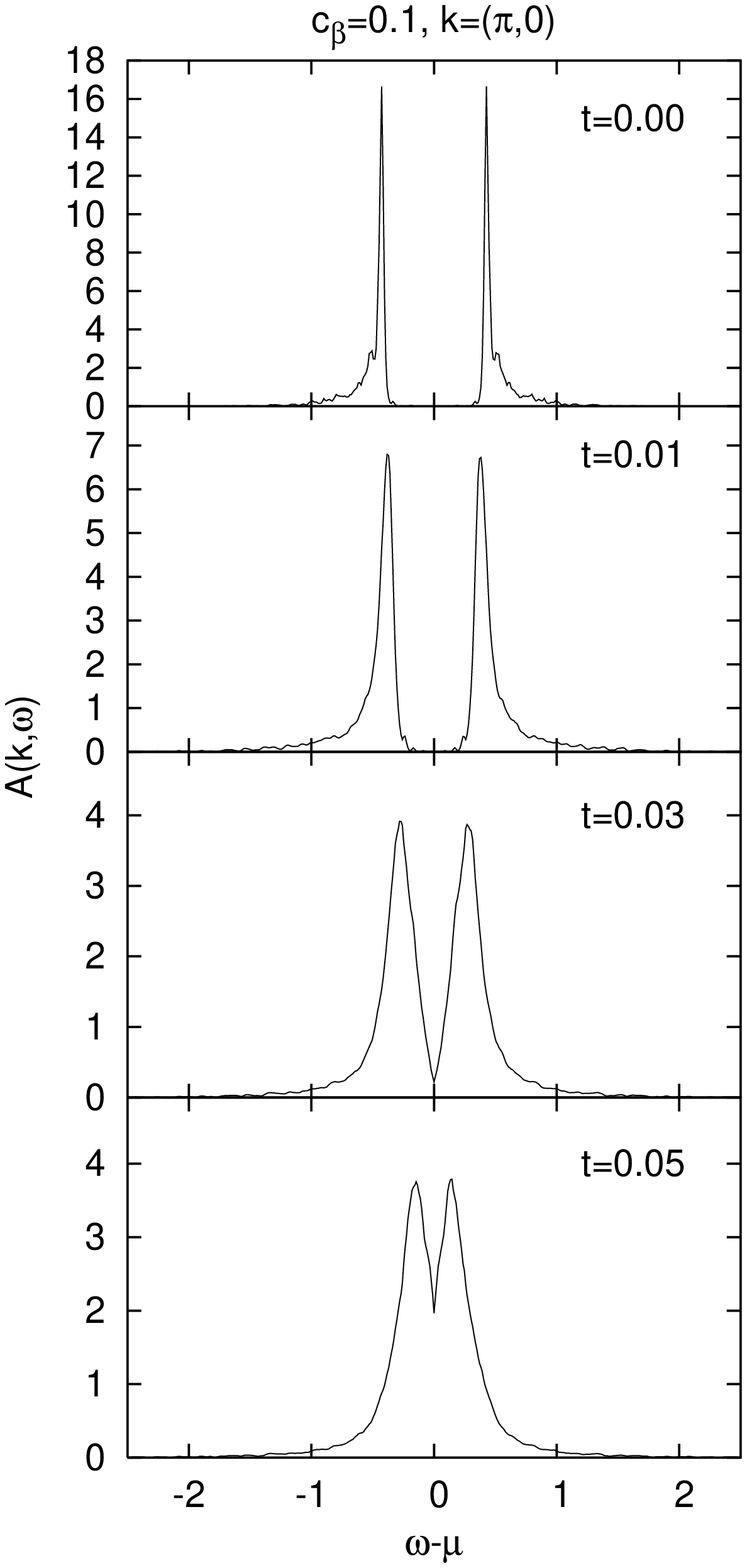}}
  \caption{
    Spectral function $\text {A} [\mathbf k = (\pi,0), \omega]$ as a
    function of $\omega$ for a system with $c_{\beta} = 0.1$ at
    different temperatures. The phase-ordering temperature of the system is
    $t_{c}\simeq 0.035$.}
  \label{fig:multiplot_spect_funct_allT_c0.1}
\end{figure}

Fig.~\ref{fig:mixed_finite_temp} shows that, as in the
case of quenched disorder, $A({\bf k}, \omega)$ is most strongly
affected by thermal fluctuations near $\mathbf k = (\pi,0)$, where it
broadens more and more with increasing $T$.  In addition to this
broadening, the peaks can be seen to shift towards smaller energies.
This behavior can be seen more clearly in
Fig.~\ref{fig:multiplot_spect_funct_allT_c0.1}, which shows $\text {A}
(\mathbf k, \omega)$ at $\mathbf k = (\pi,0)$ as a function of $\omega
- \mu$. We observe that at $T = 0$, $\text {A} (\mathbf k, \omega)$
shows relatively sharp peaks at $\omega - \mu = \pm 0.42$, with some
disorder-induced broadening only in the wings of the peak.  At
$t=0.01$, the peak height of $A(\mathbf k, \omega)$ decreases from
$\sim 17$ to about $\sim 7$, with a correspondingly increased width.

As the temperature is increased, the system eventually undergoes a
phase-disordering transition, above which the superconductor loses phase
coherence.  For the parameters used in Fig.~\ref{fig:multiplot_spect_funct_allT_c0.1},
this transition occurs at $t_c \simeq 0.035$.  $t_c$  
is the phase ordering transition temperature in units of $t_{hop}/k_B$.
% where $E_0$ is an
%energy scale which we have arbitrarily chosen as $t_{hop}$.  
%REVISE: I removed the definition of E_0 altogether since I don't think we need it.
%REVISE: Daniel, I couldn't find where we defined t or t_c so I added a definition.
%But I am not sure that I have done so correctly.  Have I said it right?
%DR. STROUD2: yes
We use a dimensionless temperature $t = k_BT/t_{hop}$ in these plots.  
At $t=0.03$, near but slightly below the phase ordering temperature
$t_{c}\simeq 0.035$, the height of the peak is further decreased, its
width further increased, and its energy shifted to a still lower
energy.  At $t=0.05>t_{c}$, the peak shifts still further toward lower
energy, but the maximum remains at finite energy.

\section{Discussion}

We have presented a simple model to study how the spectral function of
a model d-wave superconductor is affected by quenched inhomogeneities
and by thermal fluctuations of the superconducting order parameter.
The model consists of a BCS Hamiltonian for an order parameter with
$d_{x^{2}-y^{2}}$-wave symmetry, which has a position-dependent
pairing field, and which also undergoes finite-temperature thermal
fluctuations.  The spatial dependence we assume for the pairing field
is motivated by recent STM experiments on Bi2212:
%DANIEL, isn't it usually written BSCCO, not BISCO?
%DR. STROUD: I think so. But I think we can call it what we called it in our
%previous article instead: Bi2212, which is also widely used in the litterature
we assume two types of regions: and $\alpha$ region with a small gap,
and a $\beta$ region with a large gap.
%DANIEL: you wrote that both alpha and beta regions have small gaps.
%DR. STROUD: thanks for the correction.
To treat thermal fluctuations (of both amplitude and phase of the
superconducting order parameter), we assume that they are governed by
a suitable Ginzburg-Landau free energy functions, which we treat by
classical Monte Carlo simulations.

At $T = 0$, we find that $\text {A} (\mathbf k, \omega)$ is most
strongly affected by disorder near $\mathbf k = (\pi, 0)$.  In
general, this effect consists of a broadening of the peaks of $\text
{A} (\mathbf k, \omega )$ (plotted as a function of $\omega$ for fixed
$\mathbf k$).  However, at area fraction $c_{\beta}=0.5$, we find that
quenched disorder can have two qualitatively different effects,
depending on the relative magnitudes of $\Delta_\alpha$ and
$\Delta_\beta$.  If the difference between $\Delta_\alpha$ and
$\Delta_\beta$ is small, $A(\mathbf k, \omega)$ has a single, broad
peak for $\mathbf k$ near $(\pi, 0)$, extending from $\sim
\Delta_{\alpha}$ to $\sim \Delta_{\beta}$.  But for a large enough
difference between $\Delta_{\alpha}$ and $\Delta_{\beta}$, the
$A(\mathbf k, \omega)$ show a characteristic ``split-band'' behavior:
instead of a wide, single peak, there are {\it two} prominent peaks,
at $\omega = \Delta_{\alpha}$ and $\omega = \Delta_{\beta}$.

Thermal fluctuations of the pairing field also have their strongest
effect on $A(\mathbf k, \omega)$ near $\mathbf k = (\pi, 0)$.  The
effect consists of a gradual broadening of the $T = 0$ peaks with
increasing temperature, and also a shifting of those peaks towards
lower energies.  However, no dramatic change is noticeable near the
phase-ordering transition.
%DANIEL, I rewrote the above paragraph.  Is my comment about 
%``no dramatic change'' correct?
%DR. STROUD: yes

%REVISE: Daniel, I have tried to revise the paragraph below in response to
%both referees' comments.
Finally, we comment on the possible connection, if any, between our results and
experiment.  In recent angle-resolved photoemission studies by Yoshida
{\it et al}\cite{yoshida_uchida}, for LSCO, it was observed that for
doping level $x = 0.03$, a second branch developed in the dispersion
relation near $\mathbf k = (\pi, 0)$.  An explanation for the presence
of these two branches has recently been suggested by Mayr {\it et
  al.} \cite{mayr_alvarez_moreo_dagotto}.  These authors showed that
the extra branch could be explained by a model with quenched disorder,
in which the material breaks up into spatially separated
superconducting and antiferromagnetic patches.  

In the present work, we find that a similar effect, with two spectral peaks, can be
produced if there are spatially distinct superconducting regions with
sufficiently different superconducting gaps.  However, we also find that
a split spectral peak can be produced only if the magnitudes of the gaps
$|\Delta_\alpha|$ and $|\Delta_\beta|$, and of their difference,
is much larger than seems physically reasonable.  Specifically, unless
one of the gaps in the bimodal distribution is around $2.5t_{hop}$, we do not obtain a
split peak in $A({\bf k}, \omega)$ at the point ${\bf k} = (\pi, 0)$. For typical values
($t_{hop} \sim 200$ meV), this would represent a $|\Delta|$ of around $0.5$eV.  Since the
average value of $|\Delta|$ in most of the cuprate superconductors is $\sim 0.05$eV, it seems
most unlikely that random spatial fluctuations in $|\Delta|$, due to quenched disorder, could produce
such a large gap locally.   
Furthermore, even with such large quenched fluctuations in the gap, we need a {\em bimodal}
gap distribution to obtain a split spectral function - equally large quenched fluctuations,
but with a continuous distribution due to quenched disorder, would probably not give rise to
a split spectral function.   Therefore, it seems very improbable  that our model could account
for the second branch in the dispersion relation reported in Ref.\ \cite{yoshida_uchida}.
However, our results should give a reasonable picture of how quenched gap inhomogeneities affect
$A({\bf k}, \omega)$ in a d-wave superconductor over a range of parameters.

%REVISE: remove last paragraph, which is probably irrelevant since the gap differences seem to
%be too large.  

%We do find, that quenched disorder in $|\Delta|$ has its strongest effect
%on $A({\bf k}, \omega)$ near the point ${\bf k} = (\pi, 0)$ (and symmetry-related points).  This finding
%is not at all surprising, since the $A({\bf k}, \omega)$ in the homogeneous case is most sensitive
%to $|\Delta|$ near these points in the Brillouin zone.  

%While our model does not
%suggest an origin for these different gaps, it is encouraging that the
%presence of these different gaps gives spectral functions which
%appear, on the basis of numerical studies, to be consistent with
%experiments.

%================================================================

\section{Acknowledgments}  
We are grateful for support through the National Science Foundation
grant DMR04-13395.  We also thank Rajdeep Sensarma for useful
conversations.  
%REVISE: added acknowledgment to Rajdeep
The computations described here were carried out
using the facilities of the Ohio Supercomputing Center, with the help
of a grant of time.

%-------------------------------------------------------------------

\newpage

%---------------- 
%DANIEL, I rewrote this caption.  Does the horizontal width of the dashes 
%in (c) and (d) have any meaning?  It is not the widths of the spectral 
%lines?  Also, for what temperature are these plots shown.
%DR. STROUD: the horizontal width of the dashes has no meaning: it is how 
% software that I used plots them (several other theoretical articles, 
% e.g., \cite{mayr_alvarez_moreo_dagotto}, also have those dashes
% when ploting spectral function). They are not the widths in 
% energy of the 
% spectral lines: widths in energy show in the vertical direction instead of
% of horizontal.
% The temperature is zero for those plots, I added that in the caption.

%---------------- 

%---------------- fact3  1.0pi_0

%---------------- disordered lattice

%---------------- 
\begin{figure}
  \setlength{\unitlength}{1.0in}
  \begin{picture}(4.5,6)%(12,7)% this pair of numbers seems to 'reserve' space. If it's too small, figure overlaps previous figure
    \put(0,2.7){
      \includegraphics*[width=2in, angle=0]{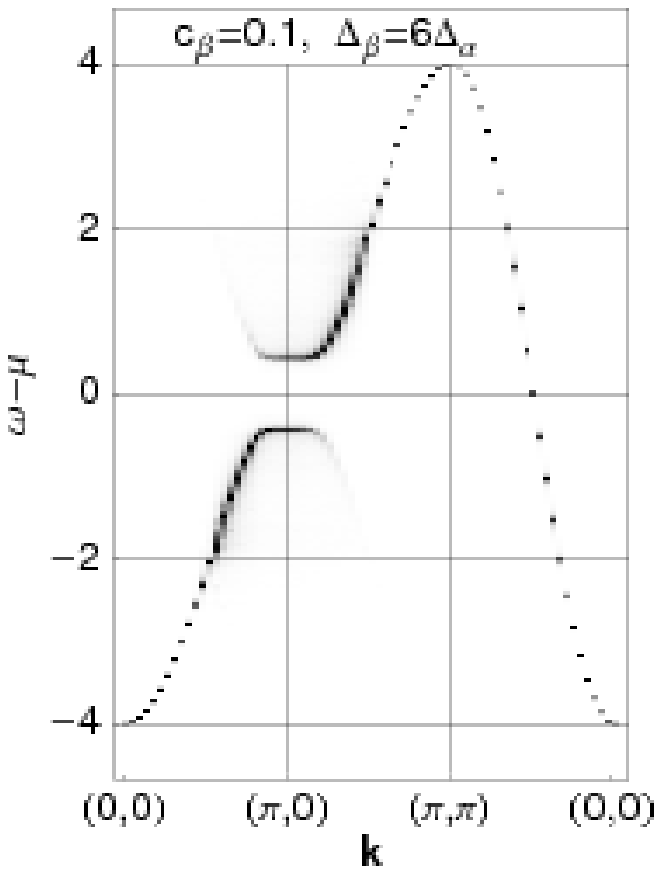}
    }
    \put(2.0,2.7){
      \includegraphics*[width=2in, angle=0]{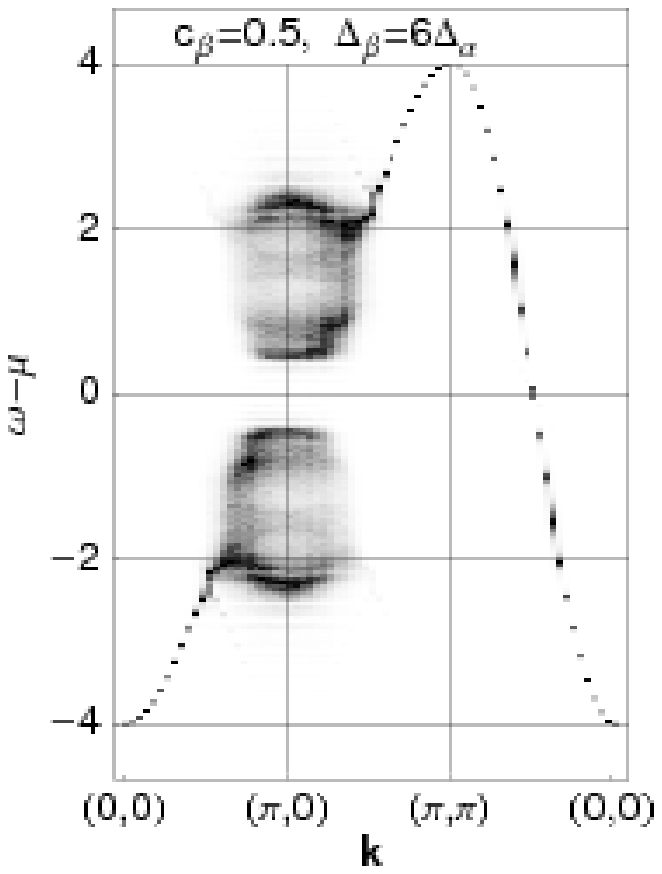}
    }
    \put(0,0){
      \includegraphics*[width=2in, angle=0]{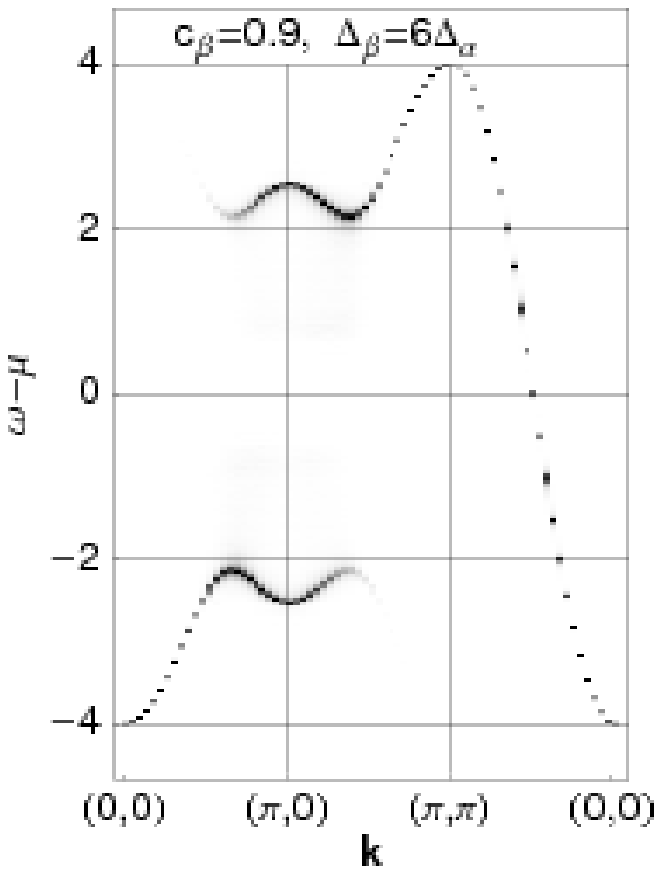}
    }
    \put(2.0,0){
      \includegraphics*[width=2in, angle=0]{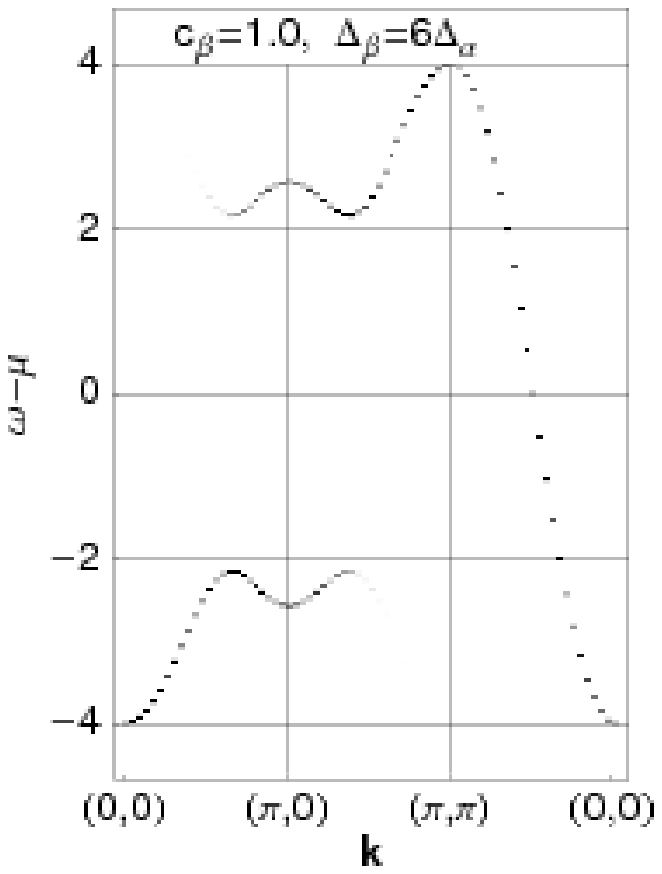}
    }
  \end{picture}
  \caption{\label{fig:mixed_fact6}
    Same as Fig.~\ref{fig:mixed_fact3}, except that $\beta$ cells have a
    $\Delta_\beta = 2.52$ instead of $\Delta_\beta = 1.26$.}
\end{figure}

%---------------- fact6  1.0pi_0
%----------------
%REVISE: Here is the proposed new Figure.  Caption has latex errors, I believe.
%Daniel, could you change the y axis label to A(k,omega) instead of A(omega, k)?  Thanks.
%DR. STROUD2: I changed all plots so that A(k,omega) instead of A(omega, k) is used.

%---------------- 
%DR. STROUD2: this is the other proposed new figure

%%---------------- fact6  1.0pi_0
%\begin{figure}[ht]\centering
%  {\includegraphics[width=10cm]{multiplot_spect_funct_fact6_L48_nb800_pi_0.ps}}
%  \caption{
%    Same as Fig.~\ref{fig:multiplot_spect_funct_fact3_1.0pi_0} but for
%    $\beta$ cells which have $\Delta = 2.52$ instead of $\Delta =
%    1.26$.}
%  \label{fig:multiplot_spect_funct_fact6_1.0pi_0}
%\end{figure}

%---------------- 

%---------------- ldos_flux0_flux1


\begin{thebibliography}{99}
\bibitem{note2} $x = 0$ is actually in the antiferromagnetic regime of LSCO. However, our model does not contain any antiferromagnetic interactions, and thus should behave similarly to $\mu$ slightly negative, which 
corresponds to strongly underdoped LSCO.
\end{thebibliography}
\end{document}